%% file: draft.tex
\documentclass[11pt]{article}
\usepackage[margin=1in]{geometry}
\usepackage[utf8]{inputenc}
\usepackage[round,numbers]{natbib}
\usepackage[dvipsnames]{xcolor}
\definecolor{linkblue}{HTML}{2484e3}
\usepackage[backref=page,colorlinks=true,citecolor=linkblue,linkcolor=linkblue]{hyperref}
\usepackage{amsmath,amsthm,amssymb,nicefrac}
\usepackage{tikz, algorithm, algpseudocode}
\usepackage{csquotes, booktabs, url, float, caption}
\usepackage{pdflscape}
\usepackage{xr}
\input{maths.tex}

\newenvironment{significance}
  {\begin{abstract}}
  {\end{abstract}}

\newcommand{\blue}[1]{{\leavevmode\color{linkblue}#1}}

\title{Fine-tuning Earth System Emulators \\ for Accurate Extreme Precipitation Projections}

\author{
  Shahine Bouabid$^{1*}$,\;
  Pedram Hassanzadeh$^{2}$,\;
  Raffaele Ferrari$^{1}$ \\[1ex]
  \small\itshape
  $^{1}$Department of Earth, Atmospheric and Planetary Sciences, \\\small\itshape Massachusetts Institute of Technology, Cambridge, USA\\
  \small\itshape
  $^{2}$Department of Geophysical Sciences, University of Chicago, Chicago, USA
}
\date{}

\begin{document}
\maketitle

\begingroup
\renewcommand{\thefootnote}{*}
\footnotetext{Corresponding author: Shahine Bouabid \url{shahineb@mit.edu}}
\endgroup







\begin{significance}
    Climate models are the basis for planning against future floods, landslides, and agricultural disruptions, but their biases misrepresent extreme rainfall and they are expensive to run long enough to sample rare events. Machine learning emulators of climate models are surrogates that can generate thousands of simulated years cheaply, but they inherit the biases from climate models. We show that briefly retraining an emulator on historical data corrects its biases in heavy rainfall projections, not just for the current climate, but also for a much warmer climate not seen in training. We establish this in a controlled experiment where a high-resolution climate model stands in for reality. We then apply the same approach to atmospheric reanalysis for the 1950-2015 period. The resulting projections suggest that rare extreme rainfall may become more frequent with warming than most current climate models suggest. This offers a new fast route for observation-informed projections of climate extremes.
\end{significance}

\begin{abstract}
    Earth System Model (ESM) simulations underpin future climate risk assessment, but available projections are limited to a few scenarios due to their computational cost, and they display known biases, particularly for precipitation extremes. ESM emulators offer inexpensive surrogates for rapidly exploring extremes in future climate scenarios, but they inherit biases from their parent ESM. Here, we show that fine-tuning an ESM emulator, trained on simulations of current and future climates, using historical data, not only corrects the biases for the current climate, but also for a much warmer climate. Because no observations of a warmer climate exist, the approach is first validated in a controlled experiment using high-resolution simulations as a proxy. Results show that fine-tuning corrects extreme precipitation biases over land, and this correction extrapolates successfully to out-of-distribution future warming scenarios. This encouraging result indicates that historical fine-tuning preserves the response of extreme precipitation to warming learned from the ESM. The tropical sensitivity of extreme precipitation is corrected, but its spatial pattern remains largely inherited from the ESM, highlighting the importance of selecting a credible ESM to train the emulator. The same recipe is then applied using observation-based historical data to project changes in extreme precipitation under future warming. With this approach, projected increases in the frequency of rare extreme precipitation events lie at the upper end of, or above, current projections. These results demonstrate that combining ESM simulations and observations within an emulator offers a new pathway to fast, observation-constrained projections of impact-relevant phenomena, with the potential to narrow projection uncertainties.
\end{abstract}

\newpage
\section*{Introduction}

Planning adaptation and mitigation measures depends on projections of future climate simulated by Earth System Models (ESMs)~\citep{flato2011earth, jones2020so}. ESM simulations, however, face two distinct limitations. First, they display systematic disagreement with observed trends and variability~\citep{shaw2024regional}, especially in the tails of the distribution of global precipitation, suggesting they may not accurately reflect what future extreme precipitation events will look like~\citep{trenberth2011changes, wehner2020characterization, abdelmoaty2021biases, john2022quantifying}. Second, the computational cost of ESM simulations makes it prohibitive to simulate the large ensembles needed to constrain the statistics of rare events. Mitigating lack of statistics and biases in the distribution tails is of paramount importance for risk assessment, because climate risk is the convolution of hazard probabilities and damages cost, which is dominated by the severity of losses occasioned by extreme events~\citep{weitzman2011fat,davenport2021contribution}. To address biases, a long line of work has used historical observations to post-process ESM precipitation projections~\citep[e.g.][]{tebaldi2007use, thackeray2022constraining, kotz2024constraining, cannon2015bias, maraun2016bias}, under the assumption that simulations that accurately reproduce the historical climate are more likely to offer reliable projections. These approaches, however, can only correct biases in events already simulated by ESMs, and do not provide a simple way to generate large ensembles of future climate realizations needed to study extremes.

Climate emulation with machine learning (ML) offers a route for generating such ensembles: once trained on climate data, ML emulators are computationally efficient for estimating the statistics of extremes because they do not explicitly resolve climate dynamics. Emulators trained on historical data can reproduce observed precipitation statistics and generate new realizations of extreme rainfall events~\citep[e.g.][]{pathak2022fourcastnet, watt2025ace2, guan2025lucie, yuval2026neural}, but their application to future precipitation has been limited because they cannot be relied on to extrapolate to warmer climates~\citep{landsberg2026forecasting, zhang2026equilibrium, merchant2026ai}. Emulators trained on ESM simulations can instead project precipitation beyond the historical warming range~\citep[e.g.][]{geogdzhayev2025eof, brenowitz2025climate, bouabid2026score}. However, they inherit the biases of the emulated ESM, including in extreme precipitation. Each class captures only part of what future precipitation projections require: fidelity to observed variability for emulators trained on historical records, and extrapolation beyond the historical climate for emulators trained on ESM simulations.

Combining observations with ESM simulations within an ML emulator may bridge this gap. An immediate strategy is transfer learning: pre-train an emulator on ESM simulations of current and future climates, then briefly finalize training on observations of the current climate to correct precipitation tails. This \emph{fine-tuning} strategy~\citep{yosinski2014transferable} rests on the expectation that patterns acquired from a comprehensive but imperfect dataset improve performance when complemented with limited but more accurate data. For temperature, evaluation against projections from a held-out model has demonstrated promise for fine-tuning on historical data~\citep{immorlano2025transferring,michel2026climate}. What is essential for extreme precipitation, and has not been addressed, is whether a correction learned at historical warming levels carries over to the forced response of extreme precipitation in warmer climates, where unprecedented events have no observational analogues. Physical scaling arguments linking warming to moisture availability and condensation rates provide a basis for expecting some predictability of extreme precipitation from temperature~\citep{pall2007testing,o2009physical,o2009scaling}. An emulator may be able to learn these physical relationships and preserve them during fine-tuning, enabling extrapolation. This is, however, hardly sufficient to build confidence that the learned relationships are consistent with physical principles governing the climate system. Evidence on the out-of-distribution skill of ML models remains mixed: they cannot generate storms more severe than those in their training data~\citep{sun2025can}, yet they are able to transfer information about extreme events across regions~\citep{sun2025predicting} and leverage short-term variability to inform longer-term trends~\citep{mansfield2020predicting,michel2026climate}.

\begin{figure}[htbp]
    \centering
    \includegraphics[width=\linewidth]{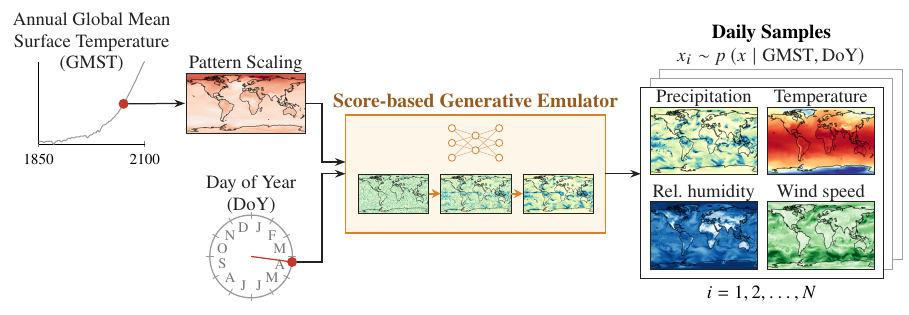}
    \caption{\textbf{Schematic of the emulator.}
    The emulator is conditioned on two inputs (red markers): the annual global mean surface temperature (GMST) and the day of year (DoY). GMST is projected to a spatial map of near-surface temperature anomaly by linear pattern scaling. DoY is encoded by a learned embedding that is concatenated with the pattern-scaled map. Given these inputs, a score-based diffusion model progressively turns Gaussian noise into a realistic global daily field (orange panel). Each run generates one joint sample $x_i$ of daily precipitation, near-surface air temperature, relative humidity and wind speed. Repeating with new noise ($i = 1, \ldots, N$) generates an ensemble of daily samples (see Methods).}
    \label{fig:1}
\end{figure}

\begin{figure}[htbp]
    \centering
    \includegraphics[width=\linewidth]{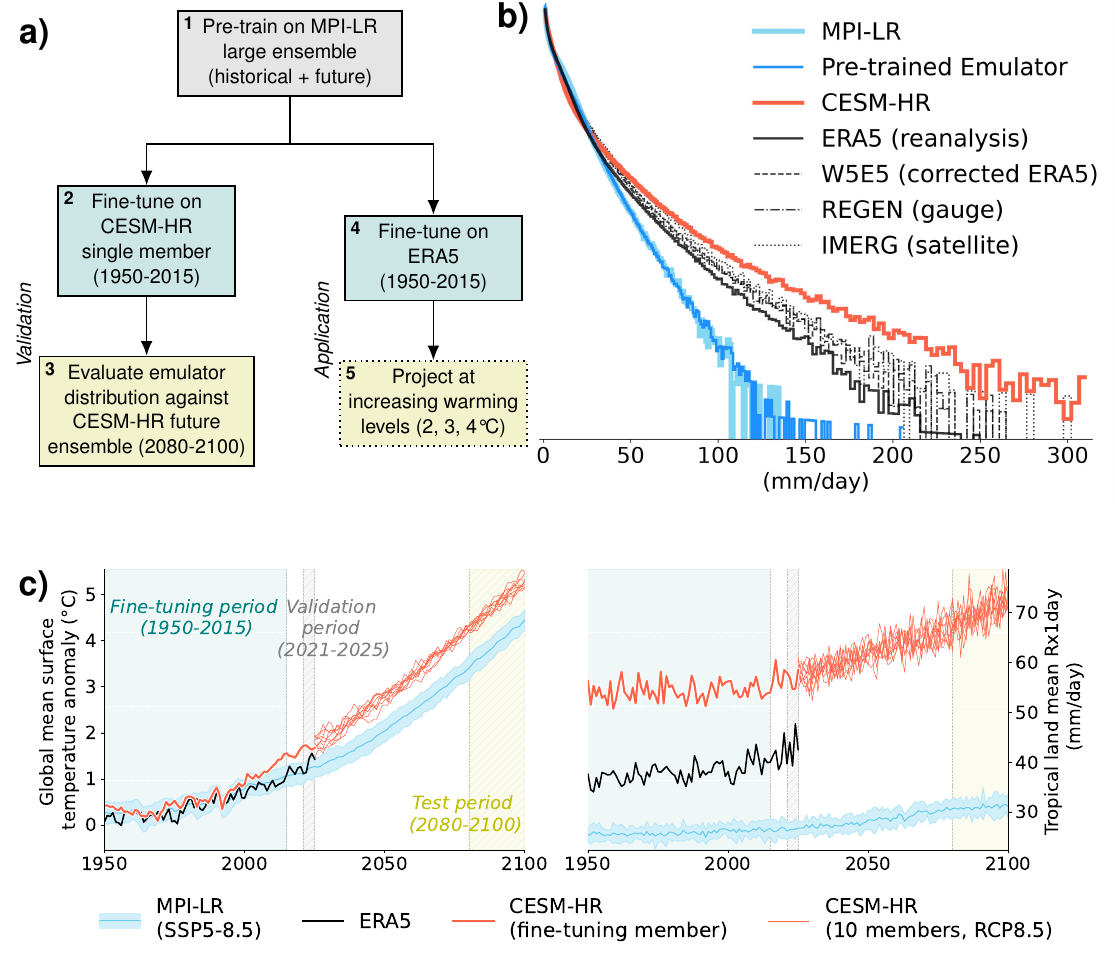}
    \caption{\textbf{Experimental setup and precipitation tail biases in ESM outputs}. (a) Schematic of the experimental design: a score-based generative emulator is pre-trained on coarse-resolution (MPI-LR) historical and future simulations, then fine-tuned on single high-resolution (CESM-HR) historical ensemble member for proof-of-concept, or on reanalysis data (ERA5) for projections. (b) Distribution of daily precipitation over land grid cells during 1979--2014 on a common 2°$\times$2° grid for the coarse-resolution model and its pre-trained emulator, the high-resolution model, reanalysis, bias-corrected reanalysis (W5E5) and two observational products (IMERG, REGEN); observations, and reanalysis products cluster together; the coarse-resolution model and its emulator underestimate precipitation tail frequencies, whereas the high-resolution model overestimates them. (c) GMST and tropical mean Rx1day trajectories for the coarse-resolution model, reanalysis, high-resolution training member and full high-resolution future ensemble; shaded regions indicate the fine-tuning period (1950-2015), validation period (2021-2025) and out-of-distribution evaluation period (2080-2100).}
    \label{fig:2}
\end{figure}

The first objective of this study is validating this approach for extreme precipitation. Because no observations of a warmer climate exist, we address this question in a controlled experiment in which a credible ground truth dataset for a future climate is available. For extreme precipitation, the lack of realism of conventional ESMs due to their coarse resolution undermines their credibility as proxies to evaluate extrapolation under warming~\citep{norris2021evaluation,donat2023credibly}. Instead, we consider an ESM emulator (Fig.~\ref{fig:1}) pre-trained on historical and future simulations from the Coupled Model Intercomparison Project (CMIP), and fine-tune it on a single historical member of a high-resolution CESM-HR ensemble~\citep{chang2020unprecedented} (Fig.~\ref{fig:2}a), which captures the spatial distribution and intensity of daily extreme precipitation more realistically than conventional ESMs~\citep{chang2026future}. We ask whether the corrections learned through historical fine-tuning carry over to future CESM-HR simulations under 8.5~Wm\textsuperscript{-2} forcing, which lie well outside historical levels. We show that fine-tuning corrects biases in the tail of the precipitation distribution from the pre-training CMIP model, and these corrections extrapolate accurately to out-of-distribution future warming, even for unprecedented events. Then, for the second objective of this study, we apply the same approach to reanalysis data (ERA5) and project changes in extreme precipitation over land under future warming.

\newpage
\section*{Results}

\subsection*{ESM extreme precipitation are biased and inherited by emulators}\label{section:2}

We pre-train the emulator on daily fields from the coarse-resolution model MPI-ESM1-2-LR~\citep{olonscheck2023new} (hereafter MPI-LR; 2°) using a score-based generative ML emulation approach~\citep{bouabid2026score}. The emulator generates joint daily fields of four near-surface variables: precipitation, temperature, relative humidity, and wind speed. Fig.~\ref{fig:2}b shows that the pre-trained emulator faithfully reproduces the tails of the precipitation distribution over land from MPI-LR, but misses the heavy precipitation events populating the distribution tail from historical reanalysis and observational products. 

Fine-tuning of the emulator is restricted to one ensemble member of the high-resolution ESM (CESM-HR~\citep{chang2020unprecedented}; 0.25°) regridded to the same 2° resolution as MPI-LR, for the 1950-2015 period. 2021-2025 is held out for hyperparameter calibration. Fine-tuning uses only 2,500 gradient steps, compared with approximately 100,000 for pre-training, and a tenfold smaller learning rate ($10^{-5}$ versus $10^{-4}$; see Methods). This \enquote{light fine-tuning} allows to avoid catastrophic forgetting~\citep{kirkpatrick2017overcoming}, where new data overwrites what has been learned during pre-training 

To assess whether CESM-HR provides a meaningful basis for model evaluation, we compare its historical land precipitation distribution to reanalysis and observational products. These products show broad mutual agreement on the magnitude of extreme precipitation. The high-resolution model has substantially heavier precipitation tail, with instead a wet bias relative to observational and reanalysis products (Fig.~\ref{fig:2}b). Matching its tail with the emulator requires a larger adjustment than matching observational products, but in the same direction. Testing whether this adjustment persists under future warming therefore provides evidence relevant to fine-tuning against observations. CESM-HR also offers improved representations of atmospheric rivers~\citep{chang2020unprecedented} and mesoscale convective systems~\citep{chang2026future} relative to coarse-resolution models, supporting its use for evaluating extreme precipitation.

We evaluate the fine-tuned emulator on against the CESM-HR 10-member ensemble under the RCP8.5 forcing scenario for the 2080-2100 period. During this period, CESM-HR exceeds its historical global mean surface temperature (GMST) levels by more than 3°C, and daily average annual maximum precipitation (hereafter Rx1day) exceed historical levels by approximately 30\%, placing the evaluation well outside the conditions seen during fine-tuning on the historical period (Fig.~\ref{fig:2}c).

\subsection*{Fine-tuning corrects precipitation tails under out-of-distribution warming}

\begin{figure}[p]
    \centering
    \includegraphics[width=\linewidth]{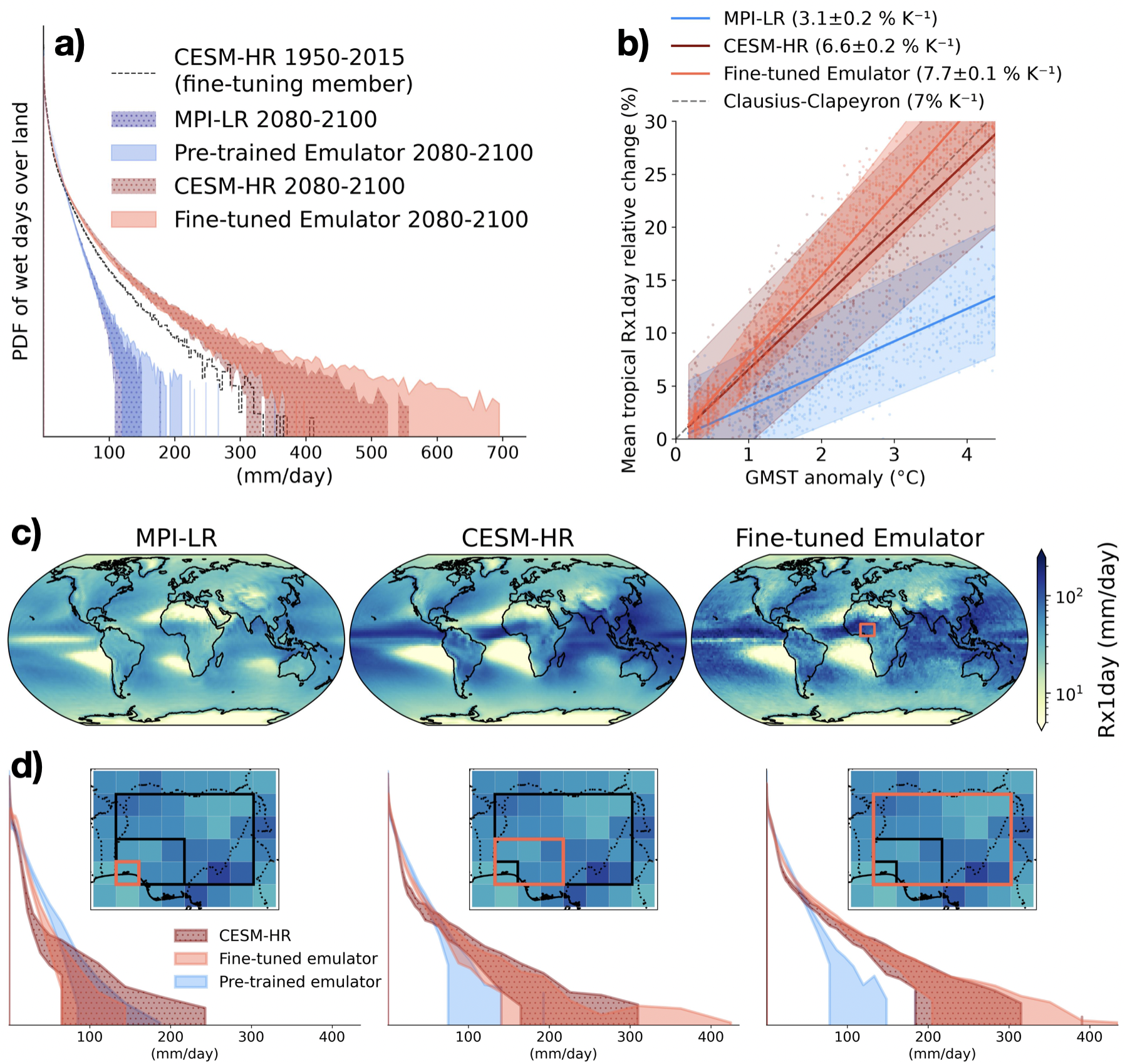}
    \caption{\textbf{Fine-tuning corrects out of sample precipitation tail biases under future warming}. (a) Distribution of precipitation over land during 2080-2100 for the coarse-resolution model (MPI-LR), high-resolution model (CESM-HR), pre-trained emulator, and fine-tuned emulator, alongside the distribution of the high-resolution historical member used for fine-tuning; the fine-tuned emulator aligns with high-resolution simulations in 2080-2100 despite being fine-tuned on only one realization of the historical period; shaded areas represent 95\% confidence range over years. For sensitivity to hyperparameters see Methods. (b) Relative change in Rx1day averaged over 30°S-30°N as a function of GMST for the coarse-resolution model, the high-resolution model and the fine-tuned emulator; the fine-tune emulator gets closer to the trend from high-resolution simulations; line indicate least-square regression, shading shows 95\% confidence range for values, and dots indicate values. (c) Maximum annual precipitation by grid-cell averaged over 2080-2100 for the coarse-resolution model, the high-resolution model, and the fine-tuned emulator; the fine-tuned emulator shows agreement with high-resolution simulations in the spatial pattern of extreme precipitation but is noisy at grid-cell level. (d) Distribution of precipitation over an increasing large region in Nigeria; the fine-tuned emulator produces distributions that align better with high-resolution simulations when pooled over larger regions.}
    \label{fig:3}
\end{figure}

After fine-tuning, the emulator shows alignment with the CESM-HR wet-day precipitation distribution over land in 2080-2100, including in the upper tail (Fig.~\ref{fig:3}a). Over the same period, the coarse-resolution ESM and the pre-trained emulator miss the most extreme precipitation events. CESM-HR simulates a significant increase in precipitation levels under warming compared to historical (SI Fig.~\blue{S1}), which is shown for the historical CESM-HR member used for fine-tuning (dashed histogram in Fig.~\ref{fig:3}a). The fine-tuned emulator reproduces this increase, providing a correction that accurately extends beyond the range of historical precipitation. Figure~\ref{fig:3}b shows that fine-tuning improves the scaling of extreme tropical precipitation with additional warming. While the coarse-resolution simulations used for pre-training show only a weak increase in Rx1day with warming (3\% K$^{-1}$), the fine-tuned emulator has a stronger sensitivity (7.7\% K$^{-1}$), closer to the CESM-HR sensitivity (6.6\% K$^{-1}$), which is more consistent with theoretical expectations (discussed later).

These results suggest a correction is learned from fine-tuning on high-resolution historical simulations, while preserving a forced response acquired from pre-training on coarse-resolution future simulations. The correction extends to temperature, relative humidity, and wind speed: fine-tuning stably shifts their end-of-century land distributions away from the coarse-resolution pre-training simulations and toward the CESM-HR future distributions (SI Fig.~\blue{S4}). Such preservation is not guaranteed as fine-tuning can distort pre-trained ML models and degrade performance outside the fine-tuning distribution~\citep{kumar2022fine}. Here, however, the emulator follows the shift simulated by CESM-HR under warming rather than reverting to, either the historical distribution used for fine-tuning, or the future distribution of the coarse-resolution simulations used for pre-training. We hypothesize that the existence of physical scaling laws relating extreme precipitation to temperature in simulations~\citep[e.g.][]{pall2007testing,o2009scaling} allows the emulator to learn during pre-training that extreme precipitation intensifies with warming. Fine-tuning then adjusts the shape of the tails and their sensitivity to warming, while preserving this learned physical tendency.

The spatial organization of future extremes precipitation for the fine-tuned emulator aligns with CESM-HR (Fig.~\ref{fig:3}c). Relative to the pre-training data (MPI-LR), the fine-tuned emulator displays stronger extremes in the tropics, a sharper change in extremes intensity across the Himalayas, and drier regions in the South Atlantic, which are consistent with CESM-HR. Emulator samples and power spectra analysis (SI Fig.~\blue{S2}, \blue{S3}) provide complementary evidence that fine-tuning modifies the emulator's spatial structure to align it with the high-resolution simulation target. Yet, Figure~\blue{S6} shows that overall large scale pattern change in Rx1day with warming remains largely inherited from the coarse-resolution pre-training simulations. It suggests that fine-tuning at historical warming levels corrects the extreme precipitation climatology, but not the spatial pattern of their sensitivity to warming. The choice of pre-training ESM thus assumes a spatial pattern for changes in extreme precipitation with warming. The emulator future distributions can nevertheless align with CESM-HR because warming-driven changes are small relative to the regional contrasts already present historically. In practice, the fine-tuned emulator can thus support regional projections where historical contrasts dominate warming-driven changes, or where the spatial pattern of change from the pre-training ESM is credible.

We also identify discrepancies attributable to the ML generative model rather than fine-tuning. Both before and after fine-tuning, the emulator occasionally produces excessively large events beyond the precipitation range observed in the training data (Figs.~\ref{fig:3}a,d). These exceedances can occur because score-based models are unreliable in low-density regions~\citep{song2019generative} and imposes no bounds on their output domain. Constraints imposed on the generative model can help mitigate these artifacts~\citep{christopher2024constrained}. Figure~\ref{fig:3}c also shows statistical artifacts in the emulated Rx1day, with a grid cell speckle that is absent from CESM-HR. The pre-trained emulator displays the same speckle, suggesting it is an issue with the generative model (SI Fig.~\blue{S7}). We estimate that emulated precipitation extremes at individual land grid cells can be under- or overestimated due to speckle by up to 10\% of their base value. Regional aggregation helps mitigate this bias: as the emulated precipitation is pooled over progressively larger regions in Nigeria, its distribution increasingly aligns with CESM-HR (Fig.~\ref{fig:3}c). Regional assessment of extreme precipitation can therefore be envisaged with the current fine-tuned emulator.

\subsection*{Future extreme precipitation fine-tuned on reanalysis}

\begin{figure}[p]
    \centering
    \includegraphics[width=\linewidth]{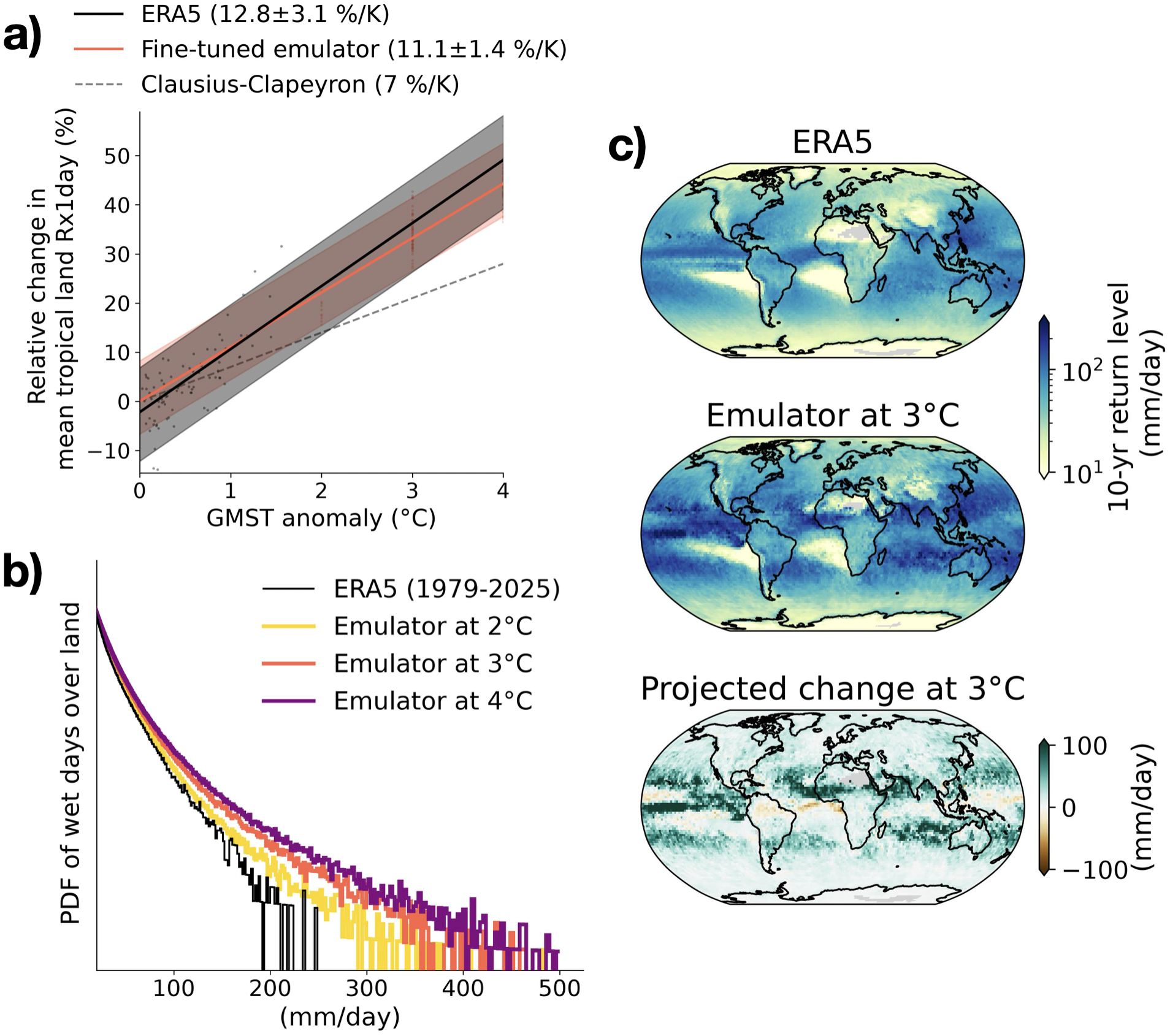}
    \caption{\textbf{Projection of extreme precipitation under increasing global warming levels}. (a) Trends in relative change of Rx1day averaged over 30°S-30°N  function of GMST, fitted for ERA5 and the emulator fine-tuned on the same data; shaded areas represent residuals 95\% confidence range. (b) Distribution of wet days over land in ERA5 and for the fine-tuned emulator at 2, 3 and 4°C above pre-industrial levels. (c) Grid-cell-level 10-year precipitation return levels estimated from ERA5 over 1979-2025, from the emulator at 3°C of warming, and the emulator projected change at 3°C taken as their difference. Gray grid-cells indicate areas with insufficient data to estimate return levels.}
    \label{fig:4}
\end{figure}

Following the evaluation against CESM-HR, we fine-tune the emulator on ERA5 reanalysis data. ERA5 tends to underrepresent the upper tail of local precipitation relative to station observations, particularly in the tropics~\citep{lavers2022evaluation}. After spatial coarsening to a horizontal resolution of 2°, it still slightly underestimates the tails of the precipitation distribution over global land, but agrees reasonably well with satellite and gauge-based observations (see Fig.~\ref{fig:2}b). We therefore proceed with ERA5 as a credible proxy for observed extreme precipitation at this resolution. We fine-tune the emulator on ERA5 data from 1950 to 2015, reserving 2021-2025 for hyperparameter tuning (see Methods), and use the resulting emulator to project global precipitation at warmings of 2°C, 3°C, and 4°C above pre-industrial levels.

The emulator's tropical Rx1day sensitivity (11.1\% K$^{-1}$) aligns with the central estimate obtained from ERA5 (Fig.~\ref{fig:4}a). Both exceed global observational estimates of approximately $6$-$10\%\,\mathrm{K}^{-1}$~\citep{sun2021global,westra2013global}. They are however consistent with evidence of higher sensitivity in the tropics; for example, \citet{o2012sensitivity} estimates a $6$-$14\%\,\mathrm{K}^{-1}$ sensitivity for the 99.9\textsuperscript{th} percentile of daily tropical precipitation. This sensitivity is reflected in a progressive shift of the global land precipitation distribution towards heavier extremes with warming (Fig.~\ref{fig:4}b). Note that because historical precipitation records are too noisy to accurately constrain sensitivities, the observational estimates here provide a broad plausibility check rather than a precise evaluation of the emulator.

The emulator's ability to generate large ensembles enables estimation of rare-event statistics relevant to climate risk. We now use this capability to examine how rainfall events with long return periods change with warming. We consider $\tau$-year return levels defined as the Rx1day that was exceeded on average once during a $\tau$ year period, a quantity often used in assessment of risk associated with extreme precipitation. Figure~\ref{fig:4}c compares the 10-year return levels estimated from ERA5 with those projected by the emulator at 3°C. It shows a general intensification of extreme precipitation over land, with the largest increases in the tropics, consistent with projections from a large majority of climate models~\citep{li2021changes}. The largest increases occur in Western and Central Africa, where return levels rise by over 100~mm/day. This regional pattern may partly reflect the response to warming in the coarse-resolution simulations used for pre-training (SI Fig.~\blue{S6}). The emulator displays plausible large-scale spatial patterns, but small-scale speckle limits quantitative evaluation of return levels at individual grid cells. We therefore restrict our analysis to regional scale using AR6 regions~\citep{iturbide2020update}.

\begin{figure}[H]
    \centering
    \includegraphics[width=0.95\linewidth]{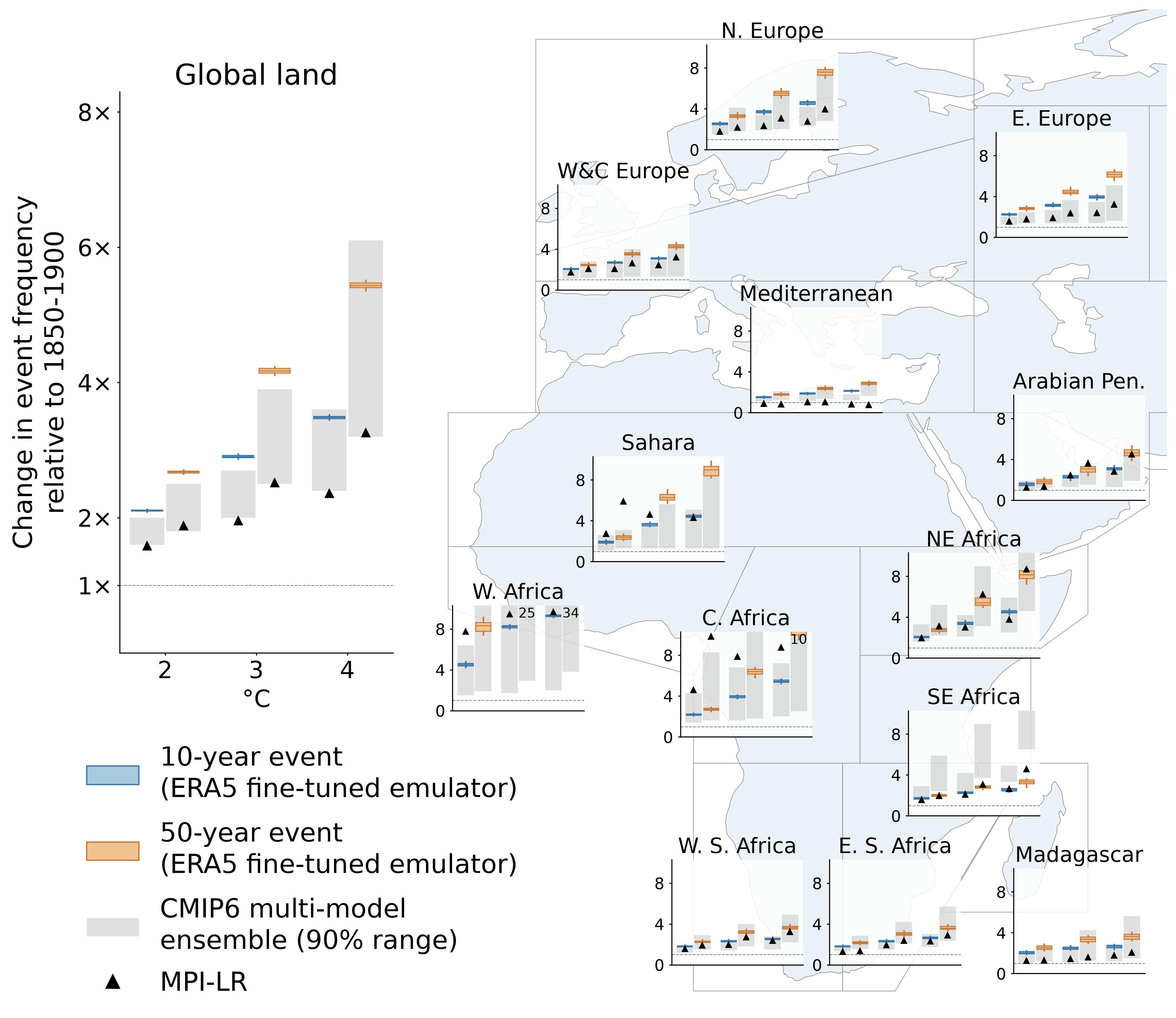}
    \caption{\textbf{Projected changes in the frequency of 10- and 50-year extreme precipitation events under 2°C, 3°C, and 4°C of global warming levels relative to pre-industrial}. Reference base period is 1850-1900. Frequency changes are expressed as multiples of the baseline frequency, with a change of $N$ indicating events occurring $N$ times as frequently as in 1850-1900. Results are shown for the global land area and for Europe, Africa and the Middle East AR6 regions (all regions shown in SI Fig.~\blue{S8}). For each box plot, the horizontal line and the box represent the median and central 66\% uncertainty range, respectively, of the frequency changes across different realizations from the emulator, and the whiskers extend to the 90\% uncertainty range. Gray shaded bars correspond to the CMIP6 multi-model ensemble 90\% range as reported in the IPCC 6\textsuperscript{th} assessment report~\citep{seneviratne2021weather}. Black triangles indicate the median equivalent change in frequency estimated from the pre-training coarse-resolution simulations (MPI-LR). Layout adapted from \citet{seneviratne2021weather}.}
    \label{fig:5}
\end{figure}

We quantify how the frequency of events with 10- and 50-year return periods in the 1850-1900 baseline period changes with warming in the emulator. Figure~\ref{fig:5} shows that, over global land, the projected increases in frequency by the emulator often lie at the upper end of, or above, the 90\% range projected by the CMIP6 multi-model ensemble, and well above the frequency ratios projected by the coarse-resolution simulations used for pre-training. The emulator projects a doubling of 10-year event frequency at 2°C of warming and a quadrupling of 50-year event frequency at 3°C. The demonstrated skill on CESM-HR suggests that fine-tuning on historical data could plausibly capture a realistic sensitivity of precipitation to warming that is stronger than CMIP6 models project. However, these increases may also reflect excessive sensitivity to warming following fine-tuning on ERA5, as the emulator lacks physical constraints on its response and may produce excessively large events.

The emulator more frequently exceeds the multi-model range at high latitudes, where heavy rainfall is generally less intense, so these frequency increases concern weaker events (SI Fig.~\blue{S8}). In tropical regions, the emulator projections lie more consistently within the multi-model range. The changes in annual frequency vary considerably among regions. In South-Western South America, a region with weaker agreement between models on the intensification of extreme rainfall~\citep{li2024constraining}, the frequency remains nearly unchanged. In contrast, in Western Africa, 50-year events become 34 times as frequent at 4°C of warming. This matches the upper bound of the CMIP6 multi-model range for the region~\citep{seneviratne2021weather}. Confidence in these projections remains limited as CMIP6 models display substantial disagreement in the region and the emulator warrants further evaluation (discussed below). Note that reported emulator uncertainty ranges are considerably narrower than multi-model ranges because the emulator is pre-trained on a single CMIP6 model and therefore does not represent inter-model variability. Rather, the ranges capture only finite-ensemble statistical estimation uncertainty arising from internal variability.

\newpage
\section*{Discussion}

We first presented a proof of concept showing that ESM emulators fine-tuned on historical data from the current climate can reproduce the forced response of extreme precipitation to warming. We then applied this approach to observation-informed historical data to propose a new methodology for regional climate projection. We first summarize the key findings to frame the discussion of their implications:

\begin{enumerate}
    \item \textbf{The distributional correction from fine-tuning persists and successfully extrapolates to out-of-distribution warming conditions:} We fine-tune an emulator, pre-trained on coarse-resolution simulations of current and future climates, on a single historical member of a high-resolution climate model that captures extreme precipitation more realistically than conventional ESMs. We then evaluate the emulator against end-of-century high-resolution simulations, well outside the historical conditions seen during fine-tuning (Fig.~\ref{fig:2}). The emulator produces joint samples of precipitation, temperature, relative humidity, and wind speed at end-of-century warming levels that align with the high-resolution simulations in both their distributions over land and their spatial characteristics (Fig.~\ref{fig:3} and Fig.~\blue{S4}). This suggests that the emulator can learn a distributional correction from its fine-tuning target while preserving the forced response learned during pre-training on a coarse-resolution ESM.
    \item \textbf{The global sensitivity of extreme precipitation to warming is set by the fine-tuning target, whereas its regional pattern is largely inherited from the pre-training simulation:} Across the tropics, the fine-tuned emulator reproduces the magnitude of change in extreme precipitation with warming in the high-resolution simulations (Fig.~\ref{fig:3}). However, the spatial pattern of change closely resembles that in the coarse-resolution pre-training data (Fig.~\blue{S6}). This suggests that fine-tuning on historical records adjusts the overall sensitivity to warming without substantially altering the spatial pattern learned from ESM simulations. Fine-tuning emulators pre-trained on different ESMs can help quantify the inter-model variability associated with the inherited response patterns. Restricting this set to ESMs whose extreme precipitation patterns satisfy constraints derived from historical observations could further improve regional projections  \citep[e.g.][]{zhang2022constraining}.
    \item \textbf{Fine-tuning on reanalysis projects future extreme precipitation at the upper-end of, or above, the CMIP6 ensemble range:} Historical biases in extreme precipitation undermine confidence in CMIP6 projections. Fine-tuning on ERA5 yields an emulator with plausible sensitivity of extreme precipitation to warming (Fig.~\ref{fig:4}). Across AR6 regions, projected changes in the frequency of extreme precipitation events lie at the upper end of the CMIP6 ensemble range, exceeding it at high latitudes and globally over land (Fig.~\ref{fig:5}). More work is needed to establish confidence in these new projections. However, they constitute plausible evidence from a validated model that data-driven approaches could narrow down the uncertainty in future extreme precipitation projections toward a greater sensitivity to warming than the CMIP6 range suggests.
\end{enumerate}

These points have important implications for projecting future precipitation extremes with ML emulators, which we discuss below. Before doing so, we emphasize that this study uses a single high-resolution model for evaluation. We selected it because it was the only model that provided a large ensemble of future simulations under strongly out-of-distribution warming and had documented, validated skill in simulating more realistic extreme precipitation than conventional ESMs~\citep{chang2026future}. It however remains a model, with its own extreme precipitation biases, and growing volumes of mesoscale and sub-mesoscale simulations produced by different modeling centers~\citep[e.g.][]{hohenegger2022icon, rackow2025multi} will offer opportunities to further test points (1) and (2). Additional simulations will help assess extrapolation skill across different physical parameterizations in models, identify new failures of the fine-tuning procedure, and provide a broader basis for interpreting projections of the fine-tuned emulator. These evaluations could strengthen confidence in, or invalidate, point (3). We also emphasize that we considered a single, relatively lightweight ML architecture (15 million parameters), which made retraining, fine-tuning, and hyperparameter searches feasible within our limited computational resources. Because generative ML models share comparable training objectives and procedures, we speculate that result (1) and limitation (2) are not specific to our ML model, but may also hold for more sophisticated architectures, which are more computationally demanding ($>$100 million parameters)~\citep[e.g.][]{brenowitz2025climate, watt2025ace2}.

We only discussed skill in capturing the tails of the distribution of daily precipitation over land. Our validation procedure selects hyperparameters for their ability to calibrate these tails, and need not be optimal for other statistics. Whether points (1) and (2) extend to other distributional features of climate remains to be tested through similar controlled experiments. Our evaluation against CESM-HR shows that fine-tuning on historical data generally improves agreement in the distributions and spatial patterns of temperature, humidity, and wind speed, including their tails under out-of-distribution future warming (Fig.~\blue{S4}). We therefore speculate that fine-tuning on historical data could also bring the spatial and temporal variability of phenomena such as heat extremes, dry spells, and strong winds into closer agreement with observations at historical warming levels, and retain skill under future warming. Extrapolation skill may depend on how well the forced response is constrained by temperature. For extreme precipitation, the Clausius–Clapeyron scaling of atmospheric moisture with temperature provides a first-order theoretical expectation of intensification at approximately 6–7\% K\textsuperscript{-1}, which can help constrain the forced response~\citep{pall2007testing,pfahl2017understanding}. Skill may also depend on how clearly the forced response emerges from internal variability in historical data. Changes in extreme precipitation emerge more robustly globally than regionally~\citep{fischer2014detection,de2023robust}. This can help understand why historical fine-tuning reproduces well future precipitation extremes at large scales (result (1)), while leaving regional response patterns largely inherited from the pre-training ESM (limitation (2)): noisy regional signals are insufficient to guide corrections. As such, one might expect greater skill in correcting regional responses for heat extremes~\citep[e.g.][]{immorlano2026guiding}, which scale well with GMST and are already emerging across many regions~\citep{seneviratne2016allowable,ranasinghe2021climate}, than for changes in circulation, for which emergence is less clearly established~\citep{ranasinghe2021climate}.

\strut
\newline
The main potential implications of points (1)-(3) for ML emulators and extreme precipitation projections are:
\begin{enumerate}
    \item[(\textit{i})]\textbf{Historical fine-tuning can preserve extrapolation to unseen warming levels.} It is surprising and encouraging that fine-tuning at historical warming levels does not destroy the ability to extrapolate extreme precipitation under warming learned from the ESM. This suggests that ML emulators may learn an internal representation of the physical response mechanism of extreme precipitation to warming, that is separate from the features modified during fine-tuning. Its restricted set of physically related input and output variables may limit opportunities to exploit spurious correlations~\citep{iglesias2024causally}, helping it learn a physically meaningful forced response that survives fine-tuning. Mechanistic interpretability approaches~\citep{macmillan2025towards} can help diagnose and understand how ESM emulators internally achieves this, i.e.\ what features of the neural network gets preserved and which one gets corrected throughout fine-tuning.
    \item[(\textit{ii})]\textbf{Local trends can remain strongly influenced by the ESM.} Although planetary scale trends extrapolate correctly, the spatial pattern of extreme precipitation trends remains inherited from the pre-training ESM. We hypothesize that the spatial scale at which fine-tuning can correct trends may be related to the scale at which historical data must be aggregated for a climate change signal to emerge. Because precipitation has much greater variability than temperature, repeating this experiment for temperature may yield improved correction of regional trends. For extreme precipitation, however, the limited emergence of regional trends in historical data warrants careful selection of credible pre-training data.
    \item[(\textit{iii})]\textbf{Fine-tuning could narrow uncertainty in regional future extreme precipitation toward the upper end of current projections.}  In our proof-of-concept, pre-training on ESM data and fine-tuning on ERA5 yields future projections that depart from those of the pre-training ESM and instead align with extrapolated ERA5 trends. A more thorough investigation is needed, particularly using different ESMs for pre-training to sample the range of responses from different models. Nevertheless, these results suggest that fine-tuning an ESM emulator with historical data could reduce uncertainty in future extreme precipitation projections by constraining the projected relationship with warming to be consistent with historical records. Future efforts should refine this approach, considering observations directly instead of reanalysis.
\end{enumerate}

These potential implications are of significant importance for climate projections and risk assessment, suggesting that ML emulators can do more than just accelerate projections: additional training on observations can improve the projections themselves. Previous work demonstrated this potential for annual mean temperature fields~\citep{immorlano2025transferring,michel2026climate}. Here, we present the first demonstration that projections of extreme daily precipitation can be improved with ML emulators. We emphasize that extensive investigation would be warranted to build confidence in these projections and make them operationally useful. This will require more thorough evaluation of extrapolation against other high-resolution models, alongside improvements to the emulator itself.

Regarding the emulator, our results identify two artifacts: the emulator occasionally generates spuriously high values and produces speckle at individual grid cells, possibly reflecting related pathologies. We therefore expect quantitative assessments at the native ESM grid scale to be unreliable with the presented method, which may also overestimate the frequency of record-breaking precipitation events. This limits its use for studying extreme precipitation changes in targeted regions and their implications for local flood risk. Regularization offers a natural starting point for mitigating these issues. Thermodynamic constraints could also penalize physically unrealistic values, while regularizing the spectral properties of generated fields could help reproduce the observed fine-scale spatial copula of extreme precipitation. Spuriously high values might also be mitigated by tailoring the generative model to heavy-tailed distributions. The score-based diffusion model used here transports Gaussian samples into climate states (Fig.~\ref{fig:1}); equivalent schemes starting from a heavy-tailed prior~\citep[e.g.][]{pandey2024heavy,shariatian2025denoising} may be better suited to sampling extreme rainfall events. 

Quantifying uncertainty in the emulator's projections is also an operational imperative~\citep{haynes2023creating} to support informed decisions based on emulator outputs. Potential approaches include conformal prediction~\citep{gopakumar2026uncertainty,asch2026rigorous} and Bayesian modeling~\citep{mansfield2026epistemic}. A fully Bayesian framework could go further by treating a generative model trained on ESM simulations as a prior and sampling future climate through posterior inference conditioned on historical data. Potential methods include guidance-based sampling for diffusion models~\citep{immorlano2026guiding} or Bayesian neural networks~\citep{mansfield2026epistemic}. Pre-training may then no longer be restricted to a single ESM, allowing a more flexible spatial pattern of sensitivity to warming, drawing on different models to better match observations. Alternative strategies using two separate models, one for ESM emulation and another for bias correction~\citep{wan2024regional} or nudging~\citep{barthel2024non,wang2025gen2}, might achieve similar success.

\section*{Conclusion}

We have demonstrated the potential of combining ESM simulations with historical data in an ML emulator to improve regional climate projections, with a focus on extreme precipitation. Using an emulator pre-trained on coarse-resolution historical and future simulations, we showed that corrections learned through historical fine-tuning extrapolate to unprecedented extremes under future warming. This was demonstrated jointly for daily precipitation, temperature, relative humidity, and wind speed, in a controlled experiment using high-resolution simulations as proxies for present and future observations. When fine-tuning on historical reanalysis, we projected an increase in frequency of extreme precipitation events that lies on the upper end of, or exceeds, current projections. We emphasize that these results do not constitute basis for trust in projections, and claims of such reliability warrants evaluation against future simulations from multiple credible high-resolution models and emergent physical properties that are not explicitly targeted during training. Our results, however, demonstrate that combining complementary sources of climate data within a single data-driven model has potential to improve our assessment of future climate extremes, and the physical risk they incur.

\newpage
\section*{Materials and Methods}

\subsection*{Coarse-resolution ESM data}

We use global daily simulations from MPI-LR~\citep{olonscheck2023new} at 2° horizontal resolution (approximately 200~km). The selected variables are 2 m temperature, precipitation, 2 m relative humidity, and 10 m wind speed. The training data include the pre-industrial control and historical experiment from the DECK CMIP6 protocol~\citep{eyring2016overview}, together with the low-forcing SSP1-2.6 and high-forcing SSP5-8.5 scenarios from the ScenarioMIP protocol~\citep{o2016scenario}. Each pathway produces a distinct GMST trajectory, ranging from low-overshoot to continued high-emission scenarios, and constitutes a diverse training set of different emulation regimes. Table~\blue{S1} summarizes the length and ensemble size of each simulation used for pre-training. All variables are pre-processed to be anomalies relative to the MPI-LR pre-industrial control climatology.

\subsection*{High-resolution ESM data}

We use the same four daily variables from a 10-member ensemble of CESM-HR simulations produced by the MESACLIP project~\citep{chang2020unprecedented}. The simulations cover 1950-2005 under historical forcing and 2006-2100 under the high-forcing RCP8.5 scenario. CESM-HR has an atmospheric horizontal resolution of 0.25° (approximately 25~km), which better resolves systems that produce extreme precipitation (e.g.\ tropical cyclones, atmospheric rivers) than conventional ESMs. Conventional ESMs at 1-2° resolution rely more heavily on parameterized subgrid convection to simulate rainfall, contributing to biases in extreme precipitation \citep[e.g.][]{john2022quantifying}. For fine-tuning, we regrid CESM-HR simulations to the 2° resolution used to pre-train the emulator. We use conservative regridding for precipitation to preserve area-integrated precipitation flux and bilinear regridding for the more spatially continuous temperature, relative humidity, and wind speed fields.

\subsection*{Observation-based historical data}

We obtain hourly ERA5 reanalysis data for 1950-2025 from the Copernicus Climate Data Store and process it to derive the same four daily variables. We derive 2~m relative humidity from specific humidity, temperature, and pressure, and 10~m wind speed from the 10~m zonal and meridional wind components. We then average the hourly fields to daily values and regrid them to the 2° resolution used to pre-train the emulator.

For the land precipitation comparison in Figure~\ref{fig:2}b we also use W5E5v2~\citep{lange2021wfde5}, an ERA5-based dataset bias-corrected for use in impact assessments ; IMERG~\citep{huffman2020integrated}, a global satellite-derived precipitation product at 0.1° resolution; and REGEN~\citep{contractor2020rainfall}, a land precipitation product gridded from station measurements at 1° resolution. The comparison uses ERA5, W5E5v2, and REGEN over 1979–2014, and IMERG over 1998–2014. We conservatively regrid precipitation from all four products to a common 2° grid.

\subsection*{ESM emulator}

The emulator used here extends \citet{bouabid2026score} from monthly to daily climate fields. It uses score-based diffusion~\citep{ho2020denoising, song2020score}, a generative modeling ML technique that learns to transform samples from an isotropic Gaussian into approximate states that an ESM could have simulated, under a given a forcing. Formally, if $p_0$ denotes a pure isotropic Gaussian distribution, and $p_1$ is the ensemble distribution produced by realizations of the ESM, the emulator evolves a trajectory that transports samples $x_0, x_1$ between these two distributions following
\begin{equation}\label{eq:ode}
    \dot x_t = \frac{1}{2}g(t)^2\nabla\log p_t\left(x_t\mid \operatorname{GMST}, \text{\texttt{dayofyear}}\right), \quad 0\leq t\leq 1
\end{equation}
where $g(t)$ is a decreasing diffusion coefficient.  The gradient of the log-density $\nabla\log p_t$, known as the \emph{score function}, is learned using a neural network. It identifies a direction that nudges $x_t$ toward regions of higher density under the ESM data distribution, given a prescribed GMST anomaly and day of the year. Therefore, starting from a pure Gaussian sample $x_0\sim p_0$, we integrate Eq.~\ref{eq:ode} to generate a spatially coherent field $x_1$ whose distribution approximates $p_1$.

To adapt the monthly emulator to daily data, we condition it on day of the year rather than month. We also add a learned positional embedding, which breaks the translation invariance of the convolutional operations in the neural network and allows to process locations differently, which we empirically found improved performance. All other modeling choices, including preprocessing, the training objective, noise schedule, and the sampling procedure, follow \citet{bouabid2026score}. Once trained, the emulator takes approximately 0.1 seconds on an H100 to generate a joint sample of all four variables for a specified day of the year and GMST anomaly.

\subsection*{Pre-training}

We pre-train the emulator using the Adam optimizer~\citep{kingma2014adam} with a learning rate $\eta=10^{-4}$ and a batch size of 32 daily fields randomly sampled across experiments, members, and times. Training runs for three epochs, corresponding to approximately $N=100,000$ gradient steps. We maintain an exponential moving average of the network parameters with a decay factor of 0.999. Pre-training takes approximately six hours on an H100 GPU. \citet{bouabid2026score} showed that the original monthly emulator closely reproduces MPI-LR distributions and key forced responses under transient future warming. Because the version used here generates daily fields, we additionally evaluate its daily annual cycle. Across all AR6 regions, we find it faithfully reproduces the precipitation and temperature seasonal cycles simulated by MPI-LR in its pre-industrial control run (Figures S9 and S10).

\subsection*{Fine-tuning}

We fine-tune the emulator by resuming training after pre-training on MPI-LR, using historical data from 1950–2015. We explore the sensitivity of fine-tuning to two hyperparameters: the number of fine-tuning steps $N$, and the learning rate $\eta > 0$, which controls how much the emulator parameters change at each fine-tuning step. Following common ML fine-tuning practice~\citep[e.g.][]{immorlano2025transferring}, we explore lower learning rates and fewer training steps than during pre-training to limit changes to the emulator's learned parameters and reduce the risk of forgetting the pre-training data. We thus explore $\eta \in \{3\cdot10^{-6}, 10^{-5}, 3\cdot10^{-5}\}$ and $N \in \{500, 1000, 1500, 2000, 2500, 3000, 3500, 4000\}$. To select $(\eta, N)$, we evaluate the calibration of land precipitation tails over the held-out validation period 2021–2025 using two objectives:
\begin{align}
    Q_{>99}^{\text{under}} & = \operatorname{Mean}_{\alpha>0.99} (q_\alpha^{\text{Data}} - q_\alpha^{\text{Emu}})_+ \\
    Q_{>99}^{\text{over}} & = \operatorname{Mean}_{\alpha>0.99} (q_\alpha^{\text{Emu}} - q_\alpha^{\text{Data}})_+,
\end{align}
where $q_\alpha^{\text{Data}}$, $ q_\alpha^{\text{Emu}}$ are the $\alpha$-quantiles of the fine-tuning data and emulator land daily precipitation distributions over the validation period, respectively, and $(\cdot)_+ = \max(\cdot, 0)$. The two objectives penalize underestimation and overestimation of quantiles above the 99\textsuperscript{th} percentile.

For fine-tuning on CESM-HR simulations, we use a single member (r2i1p1f1) to reproduce the setting in which only one historical record is available. We combine that member's historical simulation for 1950-2005 with its RCP8.5 simulation for 2006-2015. Although the latter uses scenario rather than historical forcing, we expect the simulated climate over this short period to remain close to what would have been simulated under historical forcing. We calculate a climatology from the same member over 1940-1980 and express all variables as anomalies relative to it. This matches the pre-training treatment, in which variables are expressed as anomalies relative to the MPI-LR climatology. Figure~\blue{S12} shows both validation objectives across the tested learning rates and number of fine-tuning steps. $(\eta, N) = (10^{-5}, 2500)$ is selected to balance underestimation and overestimation of precipitation tail quantiles over the validation period. For fine-tuning on ERA5, we also express all variables as anomalies relative to an ERA5 climatology calculated over 1940-1980. Evaluating the same validation objectives across the tested learning rates and number of fine-tuning steps selects $(\eta, N) = (10^{-5}, 2000)$ (SI Fig.~\blue{S13}).

\subsection*{Post-processing of fine-tuned emulator samples}

The score-based diffusion framework used for the emulator can generate values anywhere on the real line, including negative precipitation, relative humidity, and wind speed. The precipitation distribution naturally contains a point mass at zero corresponding to dry days, we therefore assign negative precipitation values to this existing dry-day state by setting them to zero. Relative humidity and wind speed on the other hand are continuous variables, clipping their negative values would create an artificial concentration at zero. We instead remap them gradually to positive values by fitting a continuous piecewise-linear function to the quantile mapping between emulator outputs and historical data over the historical period (SI Fig.~\blue{S14}, \blue{S15}).

The emulator generates daily samples independently, whereas the simulated data are temporally correlated. Independent sampling gives the emulator more opportunities to draw an extreme value within a year than real climate would, potentially biasing estimates of Rx1day. At each grid cell, we estimate this bias as the difference between Rx1day computed by sampling independent days from different historical years and Rx1day from the original, temporally ordered historical data (SI Fig.~\blue{S16}, \blue{S17}). We subtract the estimated bias from emulated Rx1day in the analyses shown in Figures~\ref{fig:3}-\ref{fig:5}.

\subsection*{Out-of-distribution evaluation with high-resolution simulations}

We evaluate the emulator fine-tuned on historical CESM-HR by comparing its extreme precipitation with CESM-HR simulations at warming levels beyond those used for fine-tuning. We use daily output from all 10 CESM-HR members over 2080–2100 under RCP8.5 (GMST anomaly $>$4°C above pre-industrial). We generate a matching 10-member ensemble of daily emulator output, conditioned on the GMST trajectory and day of the year over that period. For comparison, we also use 10 MPI-LR members over 2080–2100 under SSP5-8.5 and generate the same volume of data with the pre-trained emulator under the corresponding conditions. We bootstrap across ensemble members to obtain the uncertainty ranges for the distributions in Figures~\ref{fig:3}a and \ref{fig:3}d. For the trends in Figure~\ref{fig:3}b, we generate 10 emulator members for each year from 1950 to 2100, spanning the full range of GMST anomalies along the SSP5-8.5 trajectory. We estimate the trends by ordinary least squares regression using data pooled across years and members.

\subsection*{Projection of extreme precipitation after fine-tuning on ERA5}

We use the emulator fine-tuned on ERA5 to generate 200 years of daily precipitation at 2°C, 3°C, and 4°C above pre-industrial levels. At each level, we fix the GMST anomaly and sample daily fields throughout the year. To estimate the tropical Rx1day sensitivity in Figure~\ref{fig:4}a, we use a Theil–Sen estimator for ERA5 because it is more robust to individual anomalous years in the historical record. For the much larger emulator ensemble, we use ordinary least squares to estimate the Rx1day sensitivity.

To estimate return levels, we follow the block-maxima approach of \citet{li2021changes} which underlies the estimates from the CMIP6 estimates in \citet[Fig.~11.7]{seneviratne2021weather}. Although return levels could be estimated empirically from emulator samples, this parametric approach reduces sampling variability and provides a consistent estimation method for ERA5, emulator output, and CMIP6 models. We calculate Rx1day for each year and grid cell, then pool the annual maxima within a centered 3$\times$3 grid-cell neighborhood to increase the sample available for fitting. At each grid-cell, we fit a generalized extreme value (GEV) distribution to the pooled maxima. Its cumulative distribution function is given by
\begin{equation}
    F(x) = \begin{cases}
        \exp\left(-(1 + \xi \frac{x - \mu}{\sigma})^{-1/\xi}\right), & \xi\neq0 \\[10pt]
        \exp\left(-\exp(-\frac{x - \mu}{\sigma})\right), & \xi=0,
    \end{cases}
\end{equation}
where $\mu$, $\sigma>0$ and $\xi$ are the location, scale, and shape parameters respectively. We estimate them using L-moments~\citep{hosking1990lmoments} for computational efficiency when fitting the distribution at each grid cell. At a given location, the $\tau$-year return level $z_\tau$ is the value exceeded by an annual maximum with probability $1/\tau$, which is derived by solving $1 - F(z_\tau) = 1/\tau$. For the 10-year return levels in Figure~\ref{fig:4}c, the ERA5 fit uses 47 years of data (1979–2025) and the emulator fit uses 200 years at 3°C.

To estimate changes in the frequency of extreme precipitation in Figure~\ref{fig:5}, we first define $z_\tau^{\text{ref}}$ at each grid cell as the $\tau$-year return level for 1850–1900. We generate 10 emulator members over that period, conditioned on the associated GMST anomalies, and estimate $z_\tau^{\text{ref}}$ from their annual Rx1day. At each future warming level $\delta T\in\{2,3,4\}^\circ\mathrm{C}$, we assume that the distribution of annual maxima differs from the reference distribution only by a multiplicative factor $\rho(\delta T)$. This is an assumption used in extreme-event attribution~\citep{philip2020protocol}, which makes the estimated changes in frequency robust to anomalously high values generated by the emulator at future warming levels. The change in frequency of an extreme precipitation event of magnitude $z_\tau^{\text{ref}}$ at warming $\delta T$ is computed as
\begin{equation}
    \text{FrequencyChange}(\tau, \delta T) = \tau\left[1-F_{\text{ref}}\!\left(\frac{z_\tau^{\text{ref}}}{\rho(\delta T)}\right)\right],
\end{equation}
where $F_{\text{ref}}$ is the GEV fitted at a given grid-cell for the reference 1850–1900 period. We estimate $\rho(\delta T)$ as the ratio of median annual maxima at $\delta T$ to the median over 1850-1900. We compute the frequency ratio at each grid cell and warming level, then report the area-weighted median within each AR6 region. To assess sampling uncertainty, we repeat this 100 times, each time drawing 100 members with replacement from the 200-member emulator ensemble. We apply the same frequency-ratio estimation procedure to MPI-LR simulations, using its own 1850–1900 reference period. For each target warming level of 2°C, 3°C, or 4°C, we select climate states along the SSP5-8.5 trajectory whose GMST anomalies fall within 0.5°C of that level. The CMIP6 multi-model 90\% ranges were manually transcribed from Figure 11.7 in \citet{seneviratne2021weather}.



\section*{Acknowledgments}

S.B. and R.F. are supported by Schmidt Sciences, LLC, through the Bringing Computation to the Climate Challenge (BC3), an MIT Climate Grand Challenge Project. P.H. is supported by NSF grant AGS-2531264. We acknowledge the MIT Engaging cluster supported by the Office of Research Computing and Data, and MIT Svante cluster supported by the Center for Sustainability Science and Strategy for computing resources. We thank Tiffany Shaw, Frederic Castruccio, Y. Qiang Sun, and Chao Li for helpful suggestions, discussions, and support, that have helped shape this work.



\bibliographystyle{unsrtnat}
\bibliography{references}

\end{document}

%% file: maths.tex
\def\11{{\mathbf 1}}    



%% file: references.bib
@article{kingma2014adam,
  title={Adam: A method for stochastic optimization},
  author={Kingma, Diederik P and Ba, Jimmy},
  journal={arXiv preprint arXiv:1412.6980},
  year={2014}
}

@article{iglesias2024causally,
  title={Causally-informed deep learning to improve climate models and projections},
  author={Iglesias-Suarez, Fernando and Gentine, Pierre and Solino-Fernandez, Breixo and Beucler, Tom and Pritchard, Michael and Runge, Jakob and Eyring, Veronika},
  journal={Journal of Geophysical Research: Atmospheres},
  volume={129},
  number={4},
  pages={e2023JD039202},
  year={2024},
  publisher={Wiley Online Library}
}

@article{immorlano2026guiding,
  title={Guiding future global temperature distributions with simulations and observations},
  author={Immorlano, Francesco and Tavares, Elijah and Draxler, Felix and Nathaniel, Juan and Launeau, Violette and Watson-Parris, Duncan and Smyth, Padhraic and Gentine, Pierre and Mandt, Stephan},
  journal={arXiv preprint arXiv:2611.10882},
  year={2026}
}

@article{pfahl2017understanding,
  title={Understanding the regional pattern of projected future changes in extreme precipitation},
  author={Pfahl, Stephan and O’Gorman, Paul A and Fischer, Erich M},
  journal={Nature Climate Change},
  volume={7},
  number={6},
  pages={423--427},
  year={2017},
  publisher={Nature Publishing Group UK, London}
}

@article{kirkpatrick2017overcoming,
  title={Overcoming catastrophic forgetting in neural networks},
  author={Kirkpatrick, James and Pascanu, Razvan and Rabinowitz, Neil and Veness, Joel and Desjardins, Guillaume and Rusu, Andrei A and Milan, Kieran and Quan, John and Ramalho, Tiago and Grabska-Barwinska, Agnieszka and others},
  journal={Proceedings of the national academy of sciences},
  volume={114},
  number={13},
  pages={3521--3526},
  year={2017},
  publisher={National Academy of Sciences}
}

@article{o2009scaling,
  title={Scaling of precipitation extremes over a wide range of climates simulated with an idealized GCM},
  author={O’Gorman, Paul A and Schneider, Tapio},
  journal={Journal of Climate},
  volume={22},
  number={21},
  pages={5676--5685},
  year={2009},
  publisher={American Meteorological Society}
}

@article{pall2007testing,
  title={Testing the Clausius--Clapeyron constraint on changes in extreme precipitation under CO2 warming: P. Pall et al.: Testing the Clausius--Clapeyron constraint},
  author={Pall, Pardeep and Allen, MR and Stone, D{\'a}ith{\'\i} A},
  journal={Climate Dynamics},
  volume={28},
  number={4},
  pages={351--363},
  year={2007},
  publisher={Springer}
}

@article{o2009physical,
  title={The physical basis for increases in precipitation extremes in simulations of 21st-century climate change},
  author={O'Gorman, Paul A and Schneider, Tapio},
  journal={Proceedings of the National Academy of Sciences},
  volume={106},
  number={35},
  pages={14773--14777},
  year={2009},
  publisher={National Academy of Sciences}
}

@article{davenport2021contribution,
  title={Contribution of historical precipitation change to US flood damages},
  author={Davenport, Frances V and Burke, Marshall and Diffenbaugh, Noah S},
  journal={Proceedings of the National Academy of Sciences},
  volume={118},
  number={4},
  pages={e2017524118},
  year={2021},
  publisher={National Academy of Sciences}
}

@article{weitzman2011fat,
  title={Fat-tailed uncertainty in the economics of catastrophic climate change},
  author={Weitzman, Martin L},
  journal={Review of Environmental Economics and Policy},
  year={2011},
  publisher={The University of Chicago Press}
}

@article{philip2020protocol,
  title={A protocol for probabilistic extreme event attribution analyses},
  author={Philip, Sjoukje and Kew, Sarah and van Oldenborgh, Geert Jan and Otto, Friederike and Vautard, Robert and van Der Wiel, Karin and King, Andrew and Lott, Fraser and Arrighi, Julie and Singh, Roop and others},
  journal={Advances in Statistical Climatology, Meteorology and Oceanography},
  volume={6},
  number={2},
  pages={177--203},
  year={2020},
  publisher={Copernicus Publications G{\"o}ttingen, Germany}
}

@article{hosking1990lmoments,
  title={L-moments: analysis and estimation of distributions using linear combinations of order statistics},
  author={Hosking, Jonathan RM},
  journal={Journal of the Royal Statistical Society Series B: Statistical Methodology},
  volume={52},
  number={1},
  pages={105--124},
  year={1990},
  publisher={Oxford University Press}
}

@article{michel2026climate,
  title={From climate projections to trajectory-conditioned local predictions},
  author={Michel, Simon LL and Strommen, Kristian and Christensen, Hannah M},
  journal={EGUsphere},
  volume={2026},
  pages={1--41},
  year={2026},
  publisher={Copernicus Publications G{\"o}ttingen, Germany}
}

@incollection{huffman2020integrated,
  title={Integrated multi-satellite retrievals for the global precipitation measurement (GPM) mission (IMERG)},
  author={Huffman, George J and Bolvin, David T and Braithwaite, Dan and Hsu, Kuo-Lin and Joyce, Robert J and Kidd, Christopher and Nelkin, Eric J and Sorooshian, Soroosh and Stocker, Erich F and Tan, Jackson and others},
  booktitle={Satellite precipitation measurement: Volume 1},
  pages={343--353},
  year={2020},
  publisher={Springer}
}

@article{lange2021wfde5,
  title={WFDE5 over land merged with ERA5 over the ocean (W5E5 v2. 0)},
  author={Lange, Stefan and Menz, Christoph and Gleixner, Stephanie and Cucchi, Marco and Weedon, Graham P and Amici, Alessandro and Bellouin, Nicolas and Schmied, Hannes M{\"u}ller and Hersbach, Hans and Buontempo, Carlo and others},
  year={2021},
  publisher={ISIMIP Repository}
}

@article{contractor2020rainfall,
  title={Rainfall Estimates on a Gridded Network (REGEN)--a global land-based gridded dataset of daily precipitation from 1950 to 2016},
  author={Contractor, Steefan and Donat, Markus G and Alexander, Lisa V and Ziese, Markus and Meyer-Christoffer, Anja and Schneider, Udo and Rustemeier, Elke and Becker, Andreas and Durre, Imke and Vose, Russell S},
  journal={Hydrology and Earth System Sciences},
  volume={24},
  number={2},
  pages={919--943},
  year={2020},
  publisher={Copernicus Publications G{\"o}ttingen, Germany}
}

@article{haynes2023creating,
  title={Creating and evaluating uncertainty estimates with neural networks for environmental-science applications},
  author={Haynes, Katherine and Lagerquist, Ryan and McGraw, Marie and Musgrave, Kate and Ebert-Uphoff, Imme},
  journal={Artificial Intelligence for the Earth Systems},
  volume={2},
  number={2},
  pages={220061},
  year={2023}
}

@article{mansfield2026epistemic,
  title={Epistemic and aleatoric uncertainty quantification in weather and climate models},
  author={Mansfield, Laura A and Christensen, Hannah M},
  journal={Quarterly Journal of the Royal Meteorological Society},
  pages={e70219},
  year={2026},
  publisher={Wiley Online Library}
}

@article{gopakumar2026uncertainty,
  title={Uncertainty quantification of surrogate models using conformal prediction},
  author={Gopakumar, Vignesh and Gray, Ander and Oskarsson, Joel and Zanisi, Lorenzo and Giles, Daniel and Kusner, Matt J and Pamela, Stanislas and Peter Deisenroth, Marc},
  journal={Machine Learning: Science and Technology},
  volume={7},
  number={1},
  pages={015025},
  year={2026},
  publisher={IOP Publishing}
}

@article{asch2026rigorous,
  title={Rigorous uncertainty quantification of probabilistic AI weather forecasts with conformal prediction},
  author={Asch, Anna and Rossellini, Raphael and Hassanzadeh, Pedram and Willett, Rebecca},
  journal={arXiv preprint arXiv:2606.19642},
  year={2026}
}

@inproceedings{shariatian2025denoising,
 author = {Shariatian, Dario and Simsekli, Umut and Oliviero Durmus, Alain},
 booktitle = {International Conference on Learning Representations},
 editor = {Y. Yue and A. Garg and N. Peng and F. Sha and R. Yu},
 pages = {96991--97024},
 title = {Denoising Levy Probabilistic Models},
 url = {https://proceedings.iclr.cc/paper_files/paper/2025/file/f12c9da400fec2594a05e37b0507fd00-Paper-Conference.pdf},
 volume = {2025},
 year = {2025}
}

@article{pandey2024heavy,
  title={Heavy-tailed diffusion models},
  author={Pandey, Kushagra and Pathak, Jaideep and Xu, Yilun and Mandt, Stephan and Pritchard, Michael and Vahdat, Arash and Mardani, Morteza},
  journal={arXiv preprint arXiv:2410.14171},
  year={2024}
}

@article{macmillan2025towards,
  title={Towards mechanistic understanding in a data-driven weather model: internal activations reveal interpretable physical features},
  author={MacMillan, Theodore and Ouellette, Nicholas T},
  journal={arXiv preprint arXiv:2512.24440},
  year={2025}
}

@article{rackow2025multi,
  title={Multi-year simulations at kilometre scale with the Integrated Forecasting System coupled to FESOM2. 5 and NEMOv3. 4},
  author={Rackow, Thomas and Pedruzo-Bagazgoitia, Xabier and Becker, Tobias and Milinski, Sebastian and Sandu, Irina and Aguridan, Razvan and Bechtold, Peter and Beyer, Sebastian and Bidlot, Jean and Boussetta, Souhail and others},
  journal={Geoscientific Model Development},
  volume={18},
  number={1},
  pages={33--69},
  year={2025},
  publisher={Copernicus Publications G{\"o}ttingen, Germany}
}

@article{de2023robust,
  title={Robust global detection of forced changes in mean and extreme precipitation despite observational disagreement on the magnitude of change},
  author={de Vries, Iris Elisabeth and Sippel, Sebastian and Pendergrass, Angeline Greene and Knutti, Reto},
  journal={Earth System Dynamics},
  volume={14},
  number={1},
  pages={81--100},
  year={2023},
  publisher={Copernicus Publications G{\"o}ttingen, Germany}
}

@article{fischer2014detection,
  title={Detection of spatially aggregated changes in temperature and precipitation extremes},
  author={Fischer, Erich M and Knutti, Reto},
  journal={Geophysical Research Letters},
  volume={41},
  number={2},
  pages={547--554},
  year={2014},
  publisher={Wiley Online Library}
}

@article{hohenegger2022icon,
  title={ICON-Sapphire: simulating the components of the Earth system and their interactions at kilometer and subkilometer scales},
  author={Hohenegger, Cathy and Korn, Peter and Linardakis, Leonidas and Redler, Ren{\'e} and Schnur, Reiner and Adamidis, Panagiotis and Bao, Jiawei and Bastin, Swantje and Behravesh, Milad and Bergemann, Martin and others},
  journal={Geoscientific Model Development Discussions},
  volume={2022},
  pages={1--42},
  year={2022},
  publisher={G{\"o}ttingen, Germany}
}

@article{li2024constraining,
  title={Constraining projected changes in rare intense precipitation events across global land regions},
  author={Li, Chao and Sun, Qiaohong and Wang, Jianyu and Liang, Yongxiao and Zwiers, Francis W and Zhang, Xuebin and Li, Tong},
  journal={Geophysical Research Letters},
  volume={51},
  number={3},
  pages={e2023GL105605},
  year={2024},
  publisher={Wiley Online Library}
}

@article{li2021changes,
  title={Changes in annual extremes of daily temperature and precipitation in CMIP6 models},
  author={Li, Chao and Zwiers, Francis and Zhang, Xuebin and Li, Guilong and Sun, Ying and Wehner, Michael},
  journal={Journal of Climate},
  volume={34},
  number={9},
  pages={3441--3460},
  year={2021}
}

@article{norris2021evaluation,
  title={Evaluation of the tail of the probability distribution of daily and subdaily precipitation in CMIP6 models},
  author={Norris, Jesse and Hall, Alex and Neelin, J David and Thackeray, Chad W and Chen, Di},
  journal={Journal of Climate},
  volume={34},
  number={7},
  pages={2701--2721},
  year={2021}
}

@article{donat2023credibly,
  title={How credibly do CMIP6 simulations capture historical mean and extreme precipitation changes?},
  author={Donat, Markus G and Delgado-Torres, Carlos and De Luca, Paolo and Mahmood, Rashed and Ortega, Pablo and Doblas-Reyes, Francisco J},
  journal={Geophysical Research Letters},
  volume={50},
  number={14},
  pages={e2022GL102466},
  year={2023},
  publisher={Wiley Online Library}
}

@article{merchant2026ai,
  title={How Do AI Climate Models Respond to Warming Across Climate Zones?},
  author={Merchant, Charlotte C and Kl{\"o}wer, Milan and Stanley-Clamp, Bradley and H{\"o}ver, Maren and Michel, Simon LL and Groot, Edward and Christensen, Hannah M},
  journal={arXiv preprint arXiv:2608.17986},
  year={2026}
}

@article{christopher2024constrained,
  title={Constrained synthesis with projected diffusion models},
  author={Christopher, Jacob K and Baek, Stephen and Fioretto, Ferdinando},
  journal={Advances in Neural Information Processing Systems},
  volume={37},
  pages={89307--89333},
  year={2024}
}

@article{song2019generative,
  title={Generative modeling by estimating gradients of the data distribution},
  author={Song, Yang and Ermon, Stefano},
  journal={Advances in neural information processing systems},
  volume={32},
  year={2019}
}

@article{kumar2022fine,
  title={Fine-tuning can distort pretrained features and underperform out-of-distribution},
  author={Kumar, Ananya and Raghunathan, Aditi and Jones, Robbie and Ma, Tengyu and Liang, Percy},
  journal={arXiv preprint arXiv:2202.10054},
  year={2022}
}

@article{sun2025predicting,
  title={Predicting beyond training data via extrapolation versus translocation: AI weather models and Dubai's unprecedented 2024 rainfall},
  author={Sun, Y Qiang and Hassanzadeh, Pedram and Shaw, Tiffany and Pahlavan, Hamid A},
  journal={arXiv preprint arXiv:2505.10241},
  year={2025}
}

@article{yosinski2014transferable,
  title={How transferable are features in deep neural networks?},
  author={Yosinski, Jason and Clune, Jeff and Bengio, Yoshua and Lipson, Hod},
  journal={Advances in neural information processing systems},
  volume={27},
  year={2014}
}

@article{zhang2026equilibrium,
  title={The equilibrium response of atmospheric machine-learning models to uniform sea surface temperature warming},
  author={Zhang, Bosong and Merlis, Timothy M},
  journal={Journal of Advances in Modeling Earth Systems},
  volume={18},
  number={9},
  pages={e2026MS005815},
  year={2026},
  publisher={Wiley Online Library}
}

@article{chang2020unprecedented,
  title={An unprecedented set of high-resolution earth system simulations for understanding multiscale interactions in climate variability and change},
  author={Chang, Ping and Zhang, Shaoqing and Danabasoglu, Gokhan and Yeager, Stephen G and Fu, Haohuan and Wang, Hong and Castruccio, Frederic S and Chen, Yuhu and Edwards, James and Fu, Dan and others},
  journal={Journal of Advances in Modeling Earth Systems},
  volume={12},
  number={12},
  pages={e2020MS002298},
  year={2020},
  publisher={Wiley Online Library}
}

@article{olonscheck2023new,
  title={The new Max Planck Institute grand ensemble with CMIP6 forcing and high-frequency model output},
  author={Olonscheck, Dirk and Suarez-Gutierrez, Laura and Milinski, Sebastian and Beobide-Arsuaga, Goratz and Baehr, Johanna and Fr{\"o}b, Friederike and Ilyina, Tatiana and Kadow, Christopher and Krieger, Daniel and Li, Hongmei and others},
  journal={Journal of Advances in Modeling Earth Systems},
  volume={15},
  number={10},
  pages={e2023MS003790},
  year={2023},
  publisher={Wiley Online Library}
}

@article{chang2026future,
  title={Future extreme precipitation amplified by intensified mesoscale moisture convergence},
  author={Chang, Ping and Fu, Dan and Liu, Xue and Castruccio, Frederic S and Prein, Andreas F and Danabasoglu, Gokhan and Wang, Xiaoqi and Bacmeister, Julio and Zhang, Qiuying and Rosenbloom, Nan and others},
  journal={Nature Geoscience},
  volume={19},
  number={1},
  pages={33--41},
  year={2026},
  publisher={Nature Publishing Group UK London}
}

@article{wan2024regional,
  title={Regional climate risk assessment from climate models using probabilistic machine learning},
  author={Wan, Zhong Yi and Lopez-Gomez, Ignacio and Carver, Robert and Schneider, Tapio and Anderson, John and Sha, Fei and Zepeda-N{\'u}{\~n}ez, Leonardo},
  journal={arXiv preprint arXiv:2412.08079},
  year={2024}
}

@article{landsberg2026forecasting,
  title={Forecasting the future with yesterday's climate: Temperature bias in AI weather and climate models},
  author={Landsberg, Jacob B and Barnes, Elizabeth A},
  journal={Geophysical Research Letters},
  volume={53},
  number={6},
  pages={e2025GL119740},
  year={2026},
  publisher={Wiley Online Library}
}

@article{john2022quantifying,
  title={Quantifying CMIP6 model uncertainties in extreme precipitation projections},
  author={John, Amal and Douville, Herv{\'e} and Ribes, Aur{\'e}lien and Yiou, Pascal},
  journal={Weather and Climate Extremes},
  volume={36},
  pages={100435},
  year={2022},
  publisher={Elsevier}
}

@article{wehner2020characterization,
  title={Characterization of long period return values of extreme daily temperature and precipitation in the CMIP6 models: Part 1, model evaluation},
  author={Wehner, Michael and Gleckler, Peter and Lee, Jiwoo},
  journal={Weather and Climate Extremes},
  volume={30},
  pages={100283},
  year={2020},
  publisher={Elsevier}
}

@article{guan2025lucie,
  title={LUCIE: A lightweight uncoupled climate emulator with long-term stability and physical consistency},
  author={Guan, Haiwen and Arcomano, Troy and Chattopadhyay, Ashesh and Maulik, Romit},
  journal={Journal of Advances in Modeling Earth Systems},
  volume={17},
  number={11},
  pages={e2025MS005152},
  year={2025},
  publisher={Wiley Online Library}
}

@article{pathak2022fourcastnet,
  title={Fourcastnet: A global data-driven high-resolution weather model using adaptive fourier neural operators},
  author={Pathak, Jaideep and Subramanian, Shashank and Harrington, Peter and Raja, Sanjeev and Chattopadhyay, Ashesh and Mardani, Morteza and Kurth, Thorsten and Hall, David and Li, Zongyi and Azizzadenesheli, Kamyar and others},
  journal={arXiv preprint arXiv:2202.11214},
  year={2022}
}

@article{watt2025ace2,
  title={ACE2: accurately learning subseasonal to decadal atmospheric variability and forced responses},
  author={Watt-Meyer, Oliver and Henn, Brian and McGibbon, Jeremy and Clark, Spencer K and Kwa, Anna and Perkins, W Andre and Wu, Elynn and Harris, Lucas and Bretherton, Christopher S},
  journal={npj Climate and Atmospheric Science},
  volume={8},
  number={1},
  pages={205},
  year={2025},
  publisher={Nature Publishing Group UK London}
}

@article{yuval2026neural,
  title={Neural general circulation models for modeling precipitation},
  author={Yuval, Janni and Langmore, Ian and Kochkov, Dmitrii and Hoyer, Stephan},
  journal={Science Advances},
  volume={12},
  number={2},
  pages={eadv6891},
  year={2026},
  publisher={American Association for the Advancement of Science}
}

@article{thackeray2022constraining,
  title={Constraining the increased frequency of global precipitation extremes under warming},
  author={Thackeray, Chad W and Hall, Alex and Norris, Jesse and Chen, Di},
  journal={Nature Climate Change},
  volume={12},
  number={5},
  pages={441--448},
  year={2022},
  publisher={Nature Publishing Group UK London}
}

@article{kotz2024constraining,
  title={Constraining the pattern and magnitude of projected extreme precipitation change in a multimodel ensemble},
  author={Kotz, Maximilian and Lange, Stefan and Wenz, Leonie and Levermann, Anders},
  journal={Journal of Climate},
  volume={37},
  number={1},
  pages={97--111},
  year={2024},
  publisher={American Meteorological Society}
}

@article{zhang2022constraining,
  title={Constraining extreme precipitation projections using past precipitation variability},
  author={Zhang, Wenxia and Furtado, Kalli and Zhou, Tianjun and Wu, Peili and Chen, Xiaolong},
  journal={Nature Communications},
  volume={13},
  number={1},
  pages={6319},
  year={2022},
  publisher={Nature Publishing Group UK London}
}

@article{cannon2015bias,
  title={Bias correction of GCM precipitation by quantile mapping: how well do methods preserve changes in quantiles and extremes?},
  author={Cannon, Alex J and Sobie, Stephen R and Murdock, Trevor Q},
  journal={Journal of climate},
  volume={28},
  number={17},
  pages={6938--6959},
  year={2015}
}

@article{tebaldi2007use,
  title={The use of the multi-model ensemble in probabilistic climate projections},
  author={Tebaldi, Claudia and Knutti, Reto},
  journal={Philosophical transactions of the royal society A: mathematical, physical and engineering sciences},
  volume={365},
  number={1857},
  pages={2053--2075},
  year={2007},
  publisher={The Royal Society London}
}

@article{trenberth2011changes,
  title={Changes in precipitation with climate change},
  author={Trenberth, Kevin E},
  journal={Climate research},
  volume={47},
  number={1/2},
  pages={123--138},
  year={2011},
  publisher={JSTOR}
}

@article{o2012sensitivity,
  title={Sensitivity of tropical precipitation extremes to climate change},
  author={O’Gorman, Paul A},
  journal={Nature Geoscience},
  volume={5},
  number={10},
  pages={697--700},
  year={2012},
  publisher={Nature Publishing Group UK London}
}

@article{abdelmoaty2021biases,
  title={Biases beyond the mean in CMIP6 extreme precipitation: A global investigation},
  author={Abdelmoaty, Hebatallah Mohamed and Papalexiou, Simon Michael and Rajulapati, Chandra Rupa and AghaKouchak, Amir},
  journal={Earth's Future},
  volume={9},
  number={10},
  pages={e2021EF002196},
  year={2021},
  publisher={Wiley Online Library}
}

@article{shaw2024regional,
  title={Regional climate change: consensus, discrepancies, and ways forward},
  author={Shaw, Tiffany A and Arias, Paola A and Collins, Mat and Coumou, Dim and Diedhiou, Arona and Garfinkel, Chaim I and Jain, Shipra and Roxy, Mathew Koll and Kretschmer, Marlene and Leung, L Ruby and others},
  journal={Frontiers in climate},
  volume={6},
  pages={1391634},
  year={2024},
  publisher={Frontiers Media SA}
}

@article{mansfield2020predicting,
  title={Predicting global patterns of long-term climate change from short-term simulations using machine learning},
  author={Mansfield, Laura A and Nowack, Peer J and Kasoar, Matt and Everitt, Richard G and Collins, William J and Voulgarakis, Apostolos},
  journal={npj Climate and Atmospheric Science},
  volume={3},
  number={1},
  pages={44},
  year={2020},
  publisher={Nature Publishing Group UK London}
}

@article{barthel2024non,
  title={A non-intrusive machine learning framework for debiasing long-time coarse resolution climate simulations and quantifying rare events statistics},
  author={Barthel Sorensen, Benedikt and Charalampopoulos, Alexis and Zhang, Shixuan and Harrop, BE and Leung, LR and Sapsis, Themistoklis P},
  journal={Journal of Advances in Modeling Earth Systems},
  volume={16},
  number={3},
  pages={e2023MS004122},
  year={2024},
  publisher={Wiley Online Library}
}

@article{sun2025can,
  title={Can AI weather models predict out-of-distribution gray swan tropical cyclones?},
  author={Sun, Y Qiang and Hassanzadeh, Pedram and Zand, Mohsen and Chattopadhyay, Ashesh and Weare, Jonathan and Abbot, Dorian S},
  journal={Proceedings of the National Academy of Sciences},
  volume={122},
  number={21},
  pages={e2420914122},
  year={2025},
  publisher={National Academy of Sciences}
}

@article{flato2011earth,
  title={Earth system models: an overview},
  author={Flato, Gregory M},
  journal={Wiley Interdisciplinary Reviews: Climate Change},
  volume={2},
  number={6},
  pages={783--800},
  year={2011},
  publisher={Wiley Online Library}
}

@article{jones2020so,
  title={So what is in an earth system model?},
  author={Jones, CD},
  journal={Journal of Advances in Modeling Earth Systems},
  volume={12},
  number={2},
  pages={e2019MS001967},
  year={2020},
  publisher={Wiley Online Library}
}

@article{geogdzhayev2025eof,
  title={An EOF-Based Emulator of Means and Covariances of Monthly Climate Fields},
  author={Geogdzhayev, Gosha and Souza, Andre N and Flierl, Glenn R and Ferrari, Raffaele},
  journal={EGUsphere},
  volume={2025},
  pages={1--37},
  year={2025},
  publisher={Copernicus Publications G{\"o}ttingen, Germany}
}

@article{bouabid2026score,
  title={Score-based generative emulation of impact-relevant Earth system model outputs},
  author={Bouabid, Shahine and Souza, Andre Nogueira and Ferrari, Raffaele},
  journal={Journal of Advances in Modeling Earth Systems},
  volume={18},
  number={3},
  pages={e2025MS005558},
  year={2026},
  publisher={Wiley Online Library}
}

@article{wang2025gen2,
	title        = {{GEN2: A Generative Prediction-Correction Framework for Long-time Emulations of Spatially-Resolved Climate Extremes}},
	author       = {Wang, Mengze and Sorensen, Benedikt Barthel and Sapsis, Themistoklis},
	year         = {2025},
	journal      = {arXiv preprint arXiv:2508.15196}
}

@article{eyring2016overview,
	title        = {{Overview of the Coupled Model Intercomparison Project Phase 6 (CMIP6) experimental design and organization}},
	author       = {Eyring, Veronika and Bony, Sandrine and Meehl, Gerald A and Senior, Catherine A and Stevens, Bjorn and Stouffer, Ronald J and Taylor, Karl E},
	year         = {2016},
	journal      = {Geoscientific Model Development},
	publisher    = {Copernicus GmbH},
	volume       = {9},
	number       = {5},
	pages        = {1937--1958}
}

@article{o2016scenario,
	title        = {{The scenario model intercomparison project (ScenarioMIP) for CMIP6}},
	author       = {O'Neill, Brian C and Tebaldi, Claudia and Van Vuuren, Detlef P and Eyring, Veronika and Friedlingstein, Pierre and Hurtt, George and Knutti, Reto and Kriegler, Elmar and Lamarque, Jean-Francois and Lowe, Jason and others},
	year         = {2016},
	journal      = {Geoscientific Model Development},
	publisher    = {Copernicus GmbH},
	volume       = {9},
	number       = {9},
	pages        = {3461--3482}
}

@article{iturbide2020update,
	title        = {{An update of IPCC climate reference regions for subcontinental analysis of climate model data: definition and aggregated datasets}},
	author       = {Iturbide, Maialen and Guti{\'e}rrez, Jos{\'e} M and Alves, Lincoln M and Bedia, Joaqu{\'\i}n and Cerezo-Mota, Ruth and Cimadevilla, Ezequiel and Cofi{\~n}o, Antonio S and Di Luca, Alejandro and Faria, Sergio Henrique and Gorodetskaya, Irina V and others},
	year         = {2020},
	journal      = {Earth System Science Data},
	publisher    = {Copernicus Publications G{\"o}ttingen, Germany},
	volume       = {12},
	number       = {4},
	pages        = {2959--2970}
}

@article{maraun2016bias,
	title        = {{Bias correcting climate change simulations-a critical review}},
	author       = {Maraun, Douglas},
	year         = {2016},
	journal      = {Current Climate Change Reports},
	publisher    = {Springer},
	volume       = {2},
	number       = {4},
	pages        = {211--220}
}

@article{immorlano2025transferring,
	title        = {{Transferring climate change physical knowledge}},
	author       = {Immorlano, Francesco and Eyring, Veronika and le Monnier de Gouville, Thomas and Accarino, Gabriele and Elia, Donatello and Mandt, Stephan and Aloisio, Giovanni and Gentine, Pierre},
	year         = {2025},
	journal      = {Proceedings of the National Academy of Sciences},
	publisher    = {National Academy of Sciences},
	volume       = {122},
	number       = {15},
	pages        = {e2413503122}
}

@incollection{ranasinghe2021climate,
	title        = {{Climate Change Information for Regional Impact and for Risk Assessment}},
	author       = {Ranasinghe, R. and Ruane, A.C. and Vautard, R. and Arnell, N. and Coppola, E. and Cruz, F.A. and Dessai, S. and Islam, A.S. and Rahimi, M. and Ruiz Carrascal, D. and Sillmann, J. and Sylla, M.B. and Tebaldi, C. and Wang, W. and Zaaboul, R.},
	year         = {2021},
	booktitle    = {Climate Change 2021: The Physical Science Basis. Contribution of Working Group I to the Sixth Assessment Report of the Intergovernmental Panel on Climate Change},
	publisher    = {Cambridge University Press},
	address      = {Cambridge, United Kingdom and New York, NY, USA},
	pages        = {1767--1926},
	doi          = {10.1017/9781009157896.014},
	editor       = {Masson-Delmotte, V. and Zhai, P. and Pirani, A. and Connors, S.L. and Péan, C. and Berger, S. and Caud, N. and Chen, Y. and Goldfarb, L. and Gomis, M.I. and Huang, M. and Leitzell, K. and Lonnoy, E. and Matthews, J.B.R. and Maycock, T.K. and Waterfield, T. and Yelekçi, O. and Yu, R. and Zhou, B.}
}

@article{ho2020denoising,
	title        = {{Denoising diffusion probabilistic models}},
	author       = {Ho, Jonathan and Jain, Ajay and Abbeel, Pieter},
	year         = {2020},
	journal      = {Advances in neural information processing systems},
	volume       = {33},
	pages        = {6840--6851}
}

@inproceedings{song2020score,
	title        = {{Score-based generative modeling through stochastic differential equations}},
	author       = {Song, Yang and Sohl-Dickstein, Jascha and Kingma, Diederik P and Kumar, Abhishek and Ermon, Stefano and Poole, Ben},
	year         = {2021},
	booktitle    = {International Conference on Learning Representations}
}

@article{brenowitz2025climate,
	title        = {{Climate in a bottle: Towards a generative foundation model for the kilometer-scale global atmosphere}},
	author       = {Brenowitz, Noah D and Ge, Tao and Subramaniam, Akshay and Manshausen, Peter and Gupta, Aayush and Hall, David M and Mardani, Morteza and Vahdat, Arash and Kashinath, Karthik and Pritchard, Michael S},
	year         = {2025},
	journal      = {arXiv preprint arXiv:2505.06474}
}

@article{seneviratne2016allowable,
	title        = {{Allowable CO2 emissions based on regional and impact-related climate targets}},
	author       = {Seneviratne, Sonia I and Donat, Markus G and Pitman, Andy J and Knutti, Reto and Wilby, Robert L},
	year         = {2016},
	journal      = {Nature}
}

@article{lavers2022evaluation,
  title={An evaluation of ERA5 precipitation for climate monitoring},
  author={Lavers, David A and Simmons, Adrian and Vamborg, Freja and Rodwell, Mark J},
  journal={Quarterly Journal of the Royal Meteorological Society},
  volume={148},
  number={748},
  pages={3152--3165},
  year={2022},
  publisher={Wiley Online Library}
}

@article{westra2013global,
  title={Global increasing trends in annual maximum daily precipitation},
  author={Westra, Seth and Alexander, Lisa V and Zwiers, Francis W},
  journal={Journal of climate},
  volume={26},
  number={11},
  pages={3904--3918},
  year={2013}
}

@article{sun2021global,
  title={A global, continental, and regional analysis of changes in extreme precipitation},
  author={Sun, Qiaohong and Zhang, Xuebin and Zwiers, Francis and Westra, Seth and Alexander, Lisa V},
  journal={Journal of Climate},
  volume={34},
  number={1},
  pages={243--258},
  year={2021}
}

@incollection{seneviratne2021weather,
  author    = {Seneviratne, Sonia I. and Zhang, Xuebin and Adnan, Muhammad
               and Badi, Wafae and Dereczynski, Claudine and Di Luca, Alejandro
               and Ghosh, Subimal and Iskandar, Ismail and Kossin, James
               and Lewis, Sophie and Otto, Friederike and Pinto, Izidine
               and Satoh, Masaki and Vicente-Serrano, Sergio M.
               and Wehner, Michael and Zhou, Botao},
  title     = {Weather and Climate Extreme Events in a Changing Climate},
  booktitle = {Climate Change 2021: The Physical Science Basis. Contribution
               of Working Group I to the Sixth Assessment Report of the
               Intergovernmental Panel on Climate Change},
  editor    = {Masson-Delmotte, Val{\'e}rie and Zhai, Panmao and Pirani, Anna
               and Connors, Sarah L. and P{\'e}an, Clotilde and Berger, Sophie
               and Caud, Nada and Chen, Yang and Goldfarb, Leah
               and Gomis, Melissa I. and Huang, Mengtian and Leitzell, Katherine
               and Lonnoy, Elisabeth and Matthews, J. B. Robin
               and Maycock, Tom K. and Waterfield, Tim and Yelek{\c{c}}i, {\"O}zge
               and Yu, Rong and Zhou, Botao},
  year      = {2021},
  pages     = {1513--1766},
  publisher = {Cambridge University Press},
  address   = {Cambridge, United Kingdom and New York, NY, USA},
  doi       = {10.1017/9781009157896.013},
  url       = {https://doi.org/10.1017/9781009157896.013}
}
